\documentclass[journal=apchd5, manuscript=article, layout=traditional]{achemso}

\usepackage{amsmath,amssymb}
\usepackage{graphicx}
\usepackage{float}
\usepackage{natbib}
\usepackage{color}
\usepackage[colorlinks=true,bookmarks=false,citecolor=blue]{hyperref}
\usepackage{wrapfig}

\graphicspath{ {./Figures/} }

\newcommand{\br}{\mathbf{r} }

\renewcommand{\Im}{\frak{I}\mathrm{m} }
\newcommand{\dyad}[1] {\overset\leftrightarrow{\mathbf{#1}}}

\newcommand{\dyadG}{\dyad{\mathbf{G} } }

\newcommand{\un}{\widehat{\mathbf{n}} }

\title{Single-handedness chiral optical cavities}

\author{Kirill Voronin}
\affiliation{Center for Photonics and 2D Materials, Moscow Institute of Physics and Technology, Dolgoprudny 141700, Russia}

\author{Alexey Taradin}
\affiliation{Center for Photonics and 2D Materials, Moscow Institute of Physics and Technology, Dolgoprudny 141700, Russia}

\author{Maxim V. Gorkunov}
\affiliation{Shubnikov Institute of Crystallography, FSRC ``Crystallography and Photonics'' Russian Academy of Sciences, Moscow 119333, Russia}

\author{Denis G. Baranov}
\affiliation{Center for Photonics and 2D Materials, Moscow Institute of Physics and Technology, Dolgoprudny 141700, Russia}
\alsoaffiliation{Department of Physics, Chalmers University of Technology, 412 96, Göteborg, Sweden}
\email{denis.baranov@phystech.edu}


\begin{document}

\begin{abstract}
Geometrical chirality is a universal property encountered on very different length scales ranging from geometrical shapes of living organisms to protein and DNA molecules. Interaction of chiral matter with chiral light - that is, electromagnetic field possessing a certain handedness - underlies our ability to discriminate enantiomers of chiral molecules. In this context, it is often desired to have an optical cavity that efficiently couples to only a specific (right or left) molecular enantiomer, and does not couple to the opposite one. Here, we demonstrate a single-handedness chiral optical cavity supporting only an eigenmode of a given handedness and lacking modes having the opposite one. Resonant excitation of the cavity with light of appropriate handedness enables formation of a helical standing wave with a uniform chirality density, while the light of opposite handedness does not cause any resonant effects. Furthermore, only chiral emitters of the matching handedness efficiently interact with such a chiral eigenmode, enabling the handedness-selective strength of light-matter coupling. The proposed system expands the set of tools available for investigations of chiral matter and opens the door to studies of chiral electromagnetic vacuum.

\end{abstract}

\maketitle

\section{Introduction}

A geometrical body in three-dimensional space is said to be chiral, if it cannot be superimposed upon its mirror image by rotations and translations. Two mirrored versions of an object are called its left and right enantiomers.
Chiral objects are encountered at different length scales, ranging from spiral galaxies \cite{capozziello2006spiral} down to artificial nanostructures \cite{Cecconello2017} and single molecules \cite{Barron2004}.
Not only geometrical shapes, but electromagnetic field as well can be characterized with chirality \cite{Yang2011}. Chirality of light is related to the sense of twist of the electric and magnetic field lines \cite{Lipkin1964}. More technically, chirality density $C$ is the projection of the light's spin angular momentum onto its linear momentum \cite{Bliokh2011,Fernandez-Corbaton2016}.
An arbitrary spatial distribution of a harmonic electromagnetic field satisfying Maxwell's equations in free space can be decomposed into two components of well-defined handedness, which, in the momentum basis, are represented by left and right circularly-polarized plane waves \cite{Bliokh2011}.

Interaction of chiral electromagnetic field with chiral matter results in the well known effect of circular dichroism, which underlies numerous techniques of discriminating molecular enantiomers \cite{Barron2004,Inoue2004}.
Enhancing dichroic effects is typically achieved by interfacing chiral matter with various optical resonators \cite{Govorov2010,Tang2010,Tang2011,Garcia-Guirado2018}.
For this reason it is important to understand how the eigenmodes of optical cavities relate to the field states with well-defined handedness.
Considering the most popular type of optical cavity - a Fabry-Perot (FP) resonator formed by two homogeneous metallic mirrors - one could anticipate that excitation of such a cavity with a circularly polarized light would result in the formation of a chiral standing wave. In reality, eigenmodes of an FP cavity (at normal incidence) do not have any handedness - helicity flipping of the wave travelling between the two mirrors creates a standing wave with exactly zero helicity, as sketched in Fig. 1(a).
Some cavities, such as plasmonic and dielectric nanoantennas, offer regions of enhanced chirality density even upon excitation with linearly polarized light \cite{Davis2013,Schaferling2012}, but eigenmodes of these systems still do not possess a definite handedness.
In the context, the concept of duality \cite{Fernandez-Corbaton2013} plays an important role: dual structures preserve the handedness of incident light upon scattering, thus allowing existence of eigenmodes of well-defined helicity.

This poses a fundamental quest for optical cavities supporting eigenmodes of well-defined handedness \cite{hubener2021engineering}. An additional requirement that one could impose is that the structure supports an eigenmode of a certain handedness only, but not that of the opposite one (at least in the given spectral range). The resulting  \emph{single-handedness chiral optical cavity} would couple efficiently to emitters of a certain handedness, and negligibly weak to the opposite enantiomers. The idea of such a resonator has been outlined in ref. \cite{plum2015chiral} using a pair of chiral mirrors, although the behavior of the resulting structure has not been addressed.
A progress has been made in design of helicity-preserving cavities that are based either on excitations of large in-plane momentum modes \cite{Feis2020}, or arrays of dual nanoantennas \cite{Graf2019,Ho2017,Mohammadi2019}. However, they both posses mirror or inversion symmetry and, therefore, support eigenmodes of \emph{both helicities} at the same time.

In this paper, we demonstrate and theoretically investigate a particular realization of a single-handedness cavity - an optical resonator supporting a single eigenmode of well-defined helicity, and lacking modes of the opposite helicity in the appropriate spectral range. The proposed structure is an FP cavity formed by a stack of two photonic crystal slab mirrors \cite{Semnani2020}. The handedness preserving property of the mirrors allows resonant excitation of a \emph{chiral standing wave} of a certain handedness inside the cavity, while at the same time completely transmitting waves of the opposite handedness. Furthermore, such a chiral cavity mode efficiently couples to chiral point emitters of the matching handedness, while being non-resonant for emitters of the opposite handedness. 
The proposed system expands the set of tools available for investigations of chiral matter and opens the door to studies of chiral electromagnetic vacuum.

\section{Results}
Ordinary (homogeneous bi-isotropic) mirrors, such as metallic or dielectric Bragg ones, flip handedness of light upon normal incidence, see Fig. 1(a), resulting in formation of an achiral standing wave eigenmode when two such mirrors are stacked \cite{Feis2020}.
In order to implement the desired modal spectrum of the cavity, we need to have at our disposal a reflecting structure that preserves handedness of the reflected wave, but is transparent to the opposite handedness, as sketched in Figs. 1(b,c). To that end, we utilize the geometry proposed in ref. \cite{Semnani2020}. 
The helicity-preserving mirror is a dielectric photonic crystal slab with a square lattice and its unit cell possessing only the mirror symmetry with respect to the horizontal plane $M_{xy}$ and two-fold rotational symmetry around the $z$-axis $C_{2,z}$, see Fig. 1(d). This makes the whole structure to belong to the $p2$-symmetry class in terms of the wallpaper group.

The symmetries of this structure are essential in determining its polarization response.
Lowering the rotational symmetry to or below $C_{2,z}$ enables conversion between clockwise and counter-clockwise rotating polarizations and allows us to preserve the handedness in reflection at normal incidence. As soon as the unit cell has at least three-fold rotational symmetry $C_{n,z}$, $n \ge 3$, the structure preserves the optical spin and reverses the handedness of a circularly polarized plane wave at normal incidence \cite{Menzel2010}.
At the same time, the lack of any vertical plane of mirror symmetry is what enables different response for two incident handednesses and asymmetric left-to-right and right-to-left conversion.
The presence of the horizontal mirror symmetry ensures that the mirror itself is not chiral.

\begin{figure*} 
\includegraphics[width=1\textwidth]{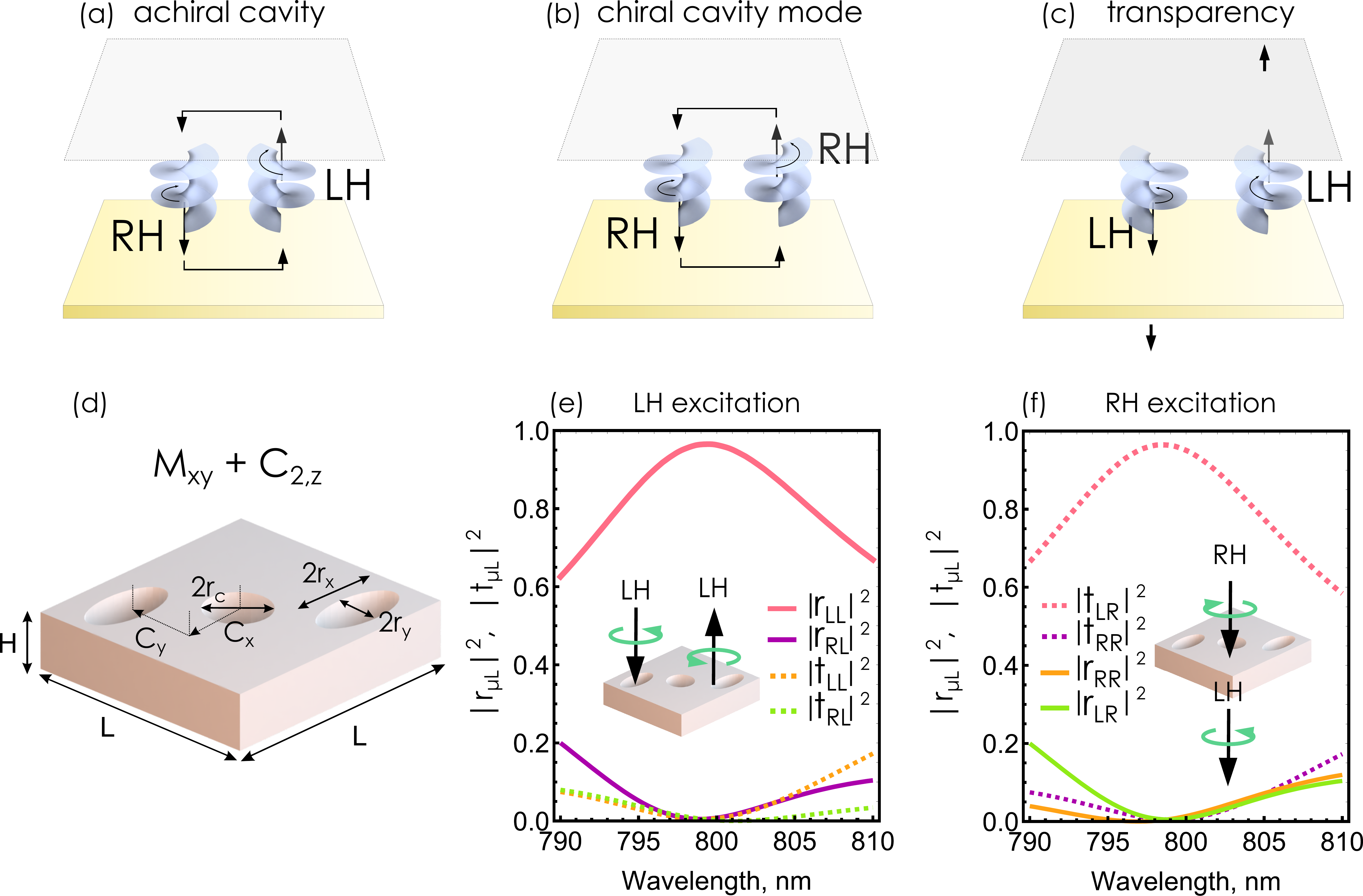}
\caption{(a-c) Concept of a single-handedness cavity. An ordinary achiral cavity, (a), does not support chiral eigenmodes due to handedness flipping of the standing wave inside the cavity. A single-handedness optical cavity supports a mode of a well-defined handedness, (b), and does not support a mode of the opposite handedness at the same wavelength, (c). (d) The unit cell of the photonic crystal slab mirror featuring handedness-preserving reflection at normal incidence. (e) Reflection and transmission coefficients of the optimized mirror in the circular polarization basis for left-handed incident light exhibiting near-unity reflection with preserved handedness. (f) Reflection and transmission coefficients of the optimized mirror for right-handed incident light exhibiting near-complete transmission accompanied by handedness flipping.}
\label{fig1}
\end{figure*}

We begin by designing a single mirror made of a dielectric with refractive index $n=4$ in air featuring the desired spectral characteristics at the target wavelength of 800 nm. 
Numerical optimization was performed with the use of particle swarm algorithm in a finite-difference time domain (FDTD) commercial solver (Lumerical).
By enforcing sub-diffraction periodicity of the mirror, we find the set of unit cell dimensions yielding near-perfect polarization conversion for normally incident light of both helicities: thickness $H = 142$~nm, unit cell size $L=366$~nm, $r_c=90$~nm, $r_x=89$~nm, $r_y=34$~nm, $C_x=59$~nm, $C_y=104$~nm. Note that this particular design results in overlapping center and corner holes, but an alternative one without overlapping holes can be found if needed.

With these dimensions, left-handed light normally incident from top is nearly perfectly reflected with handedness preservation, $|r_{LL}|^2 \approx 1$, while a right-handed incident plane wave is nearly perfectly transmitted into the opposite handedness, $|t_{LR}|^2 \approx 1$, Fig. 1(e,f).
Here $|r_{\mu \nu}|^2$ and $|t_{\mu \nu}|^2$ are power reflection and transmission coefficients of light from polarization state $\nu$ into polarization state $\mu$.
Simulated electric field and chirality density distributions clearly visualize these regimes (see Fig. S1).
In the following, we define the left-handed, LH (right-handed, RH) light in a transparent medium as a circularly-polarized plane wave with its magnetic field $\mathbf{H}(t)$ being $\pi/2$ ahead (behind) the electric field  $\mathbf{E}(t)$ ($e^{- i \omega t}$ time dependence is assumed):
\begin{align}
   \mathbf{H}_{LH}(t) = \frac{1}{Z}e^{-i \pi/2} \mathbf{E}_{LH}(t),\ \    \mathbf{H}_{RH}(t) = \frac{1}{Z}e^{+i \pi/2} \mathbf{E}_{RH}(t),
\end{align}
where $Z=\sqrt{\mu_0/\varepsilon \varepsilon_0}$ is the real-valued impedance of the medium.
Alternatively, LH (RH) light corresponds to clockwise (counter-clockwise) rotating electric and magnetic fields when viewed from the source.


The unusual polarization response of these mirrors can be understood on the basis of the chiral coupled-mode theory (CMT) \cite{kondratov2016extreme, Gorkunov2020Metasurfaces}.
The detailed description of CMT application to the photonic slab chiral mirrors is given in Supporting Information Section S2.
In this framework, the nanostructured mirror hosts a set of eigenstates described by a vector of complex amplitudes $\mathbf{p} = (p_1,...p_N)^T$. 
The structure is illuminated from both sides by a set of normally incident plane waves, which in the basis of circularly polarized waves is described by a vector of complex amplitudes $\mathbf{a} = (a_R, a_L, a_R', a_L')^T$, where primed and unprimed variables denote two different sides of the mirror.
Interaction of incident waves with the eigenstates determines the amplitudes of scattered (specular reflected and transmitted) waves, which are described by a vector $\mathbf{b} = (b_R, b_L, b_R', b_L')^T$.
The amplitudes of the incident and scattered fields are related by the $S$-matrix:
\begin{equation}\label{SLReq}
    \mathbf{b} =    \left(
    \begin{array}{cccc}
    r_{RR} & r_{RL} & t'_{RR} & t'_{RL}\\
    r_{LR} & r_{LL} & t'_{LR} & t'_{LL}\\
    t_{RR} & t_{RL} & r'_{RR} & r'_{RL}\\
    t_{LR} & t_{LL} & r'_{LR} & r'_{LL}
    \end{array}
    \right) \mathbf{a}.
\end{equation}

The dynamics of the mirror's eigenstates is described by the CMT equation:
\begin{equation}\label{COM1}
    \frac{d{\bf p}}{dt } = (i{\bf \Omega}-{\bf \Gamma}) {\bf p}+{\bf M}^T{\bf a}, 
\end{equation}
where $\mathbf{\Omega}$ contains detuning between the eigenstates and the excitation frequency, $\mathbf{\Gamma}$ contains the eigenstates' decay rates, $\mathbf{M}$ is the tensor of coupling constants between scattering channels and eigenstates.
The amplitudes $\mathbf{b}$ of the scattered field take the form:
\begin{equation}\label{COM2}
    {\bf b} = {\bf M}{\bf p} + {\bf C}{\bf a}, 
\end{equation}
where $\bf C$ describes the non-resonant scattering pathway.
For a steady state, one can resolve these equations with respect to $\bf p$ and $\bf b$ and obtain the following form of the $S$-matrix of the dielectric mirror:
\begin{equation}\label{Smatr2}
    {\bf S}={\bf C}-{\bf M}[i{\bf \Omega}-{\bf \Gamma}]^{-1}{\bf M}^T,
\end{equation}
which allows us to describe redistribution of incoming circularly polarized waves given the knowledge of the resonant frequencies and radiative coupling constants of the eigenstates of the single mirror.

The CMT analysis reveals that for our periodic mirror to exhibit the required functionality (near-perfect reflection of one polarization with handedness preservation, and transparency for the opposite one), the structure needs to support at least a pair of orthogonal eigenstates (see Supporting Information Section S2.3).
By fitting the simulated power reflection and transmission coefficients from Fig. 1(e,f) with the analytical expressions provided by the CMT, we can obtain analytical complex reflection and transmission amplitudes describing the behavior of a single helicity-preserving mirror (see Supporting Information Section S2.4 for details of the fitting procedure).

\begin{figure*}[t] 
\includegraphics[width=1\textwidth]{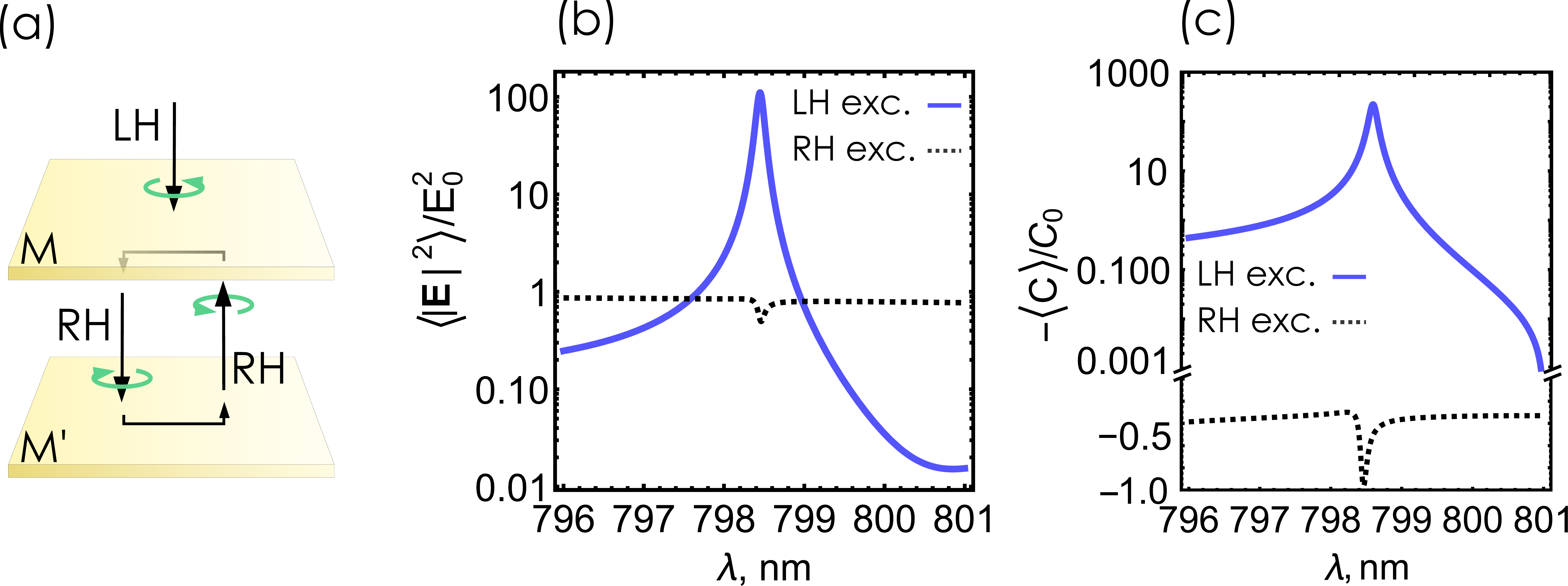}
\caption{Transfer-matrix modelling of a single-handedness cavity. (a) Illustration of the formation of a standing chiral wave in the region between two handedness-preserving mirrors. (b) Wavelength-dependent volume-average electric field intensity enhancement (calculated using the transfer-matrix approach) inside the cavity of thickness $d = 1857$ nm for RH and LH excitations at normal incidence. (c) Analogously to (b), volume-average chirality density enhancement inside the cavity (normalized by the absolute value of the chirality density of the incident circularly polarized wave $C_0 = \varepsilon_0| \mathbf{E}_{inc}|^2 / (2\omega c)$) induced by normally incident RH and LH waves.}
\label{fig2}
\end{figure*}


Consider now a stack of two such mirrors forming a cavity. 
Suppose a LH wave excites the system from the top; after transmission through the first mirror it will be converted into a RH wave. Thus for the cavity to sustain a RH chiral standing wave, the bottom mirror $M'$ needs to be a reflected (with respect to the vertical plane) version of $M$, Fig. 2(a). This should allow a RH plane wave to travel back and forth in the region between the two mirrors without ever changing its handedness.
We can use the obtained complex amplitudes and model the behavior of a cavity by applying the transfer-matrix method (see Supporting Information Section S3 for realization of the transfer-matrix method).

Fig. 2(b) presents the resulting electric field intensity enhancement averaged over the cavity interior for the center-to-center distance between the mirrors (which we will refer to in the following as the cavity thickness) $d = 1857$ nm. Clearly, incident LH field leads to a pronounced intensity enhancement, whereas excitation with RH field does not produce any.
In order to confirm that it is indeed chiral electromagnetic field that is induced by the LH incident wave, we calculate the optical chirality density $C$ inside the cavity:
\begin{equation}
    C(\mathbf{r}) = \frac{\varepsilon_0 \omega}{2} \Im [\mathbf{E} \cdot \mathbf{B}^*].
\end{equation}  
Strongly negative volume-average chirality density near the resonant cavity wavelength indicates excitation of an RH field inside the cavity, Fig. 2(c).


Next we turn to full-wave FDTD simulations of the single-handedness cavity. As we noted earlier, one mirror of the cavity should be flipped with respect to the vertical plane. This gives rise to a geometrically chiral structure lacking any mirror or inversion symmetry, Fig. 3(a).
In order to find the optimal cavity thickness sustaining a chiral eigenmode, we simulate volume-average electric field enhancement in the region inside the cavity upon normal incidence and sweep the center-to-center distance $d$ (see Fig. S9 for the sweep of the field enhancement spectra in a range of cavity thicknesses). For the cavity thickness $d = 1866$ nm the field enhancement spectra demonstrate handedness-selective electric field enhancement for LH excitation at the wavelength of $\lambda_{res} = 798.2$ nm, and much weaker enhancement for RH excitation, Fig. 3(b).

\begin{figure*} 
\includegraphics[width=1\textwidth]{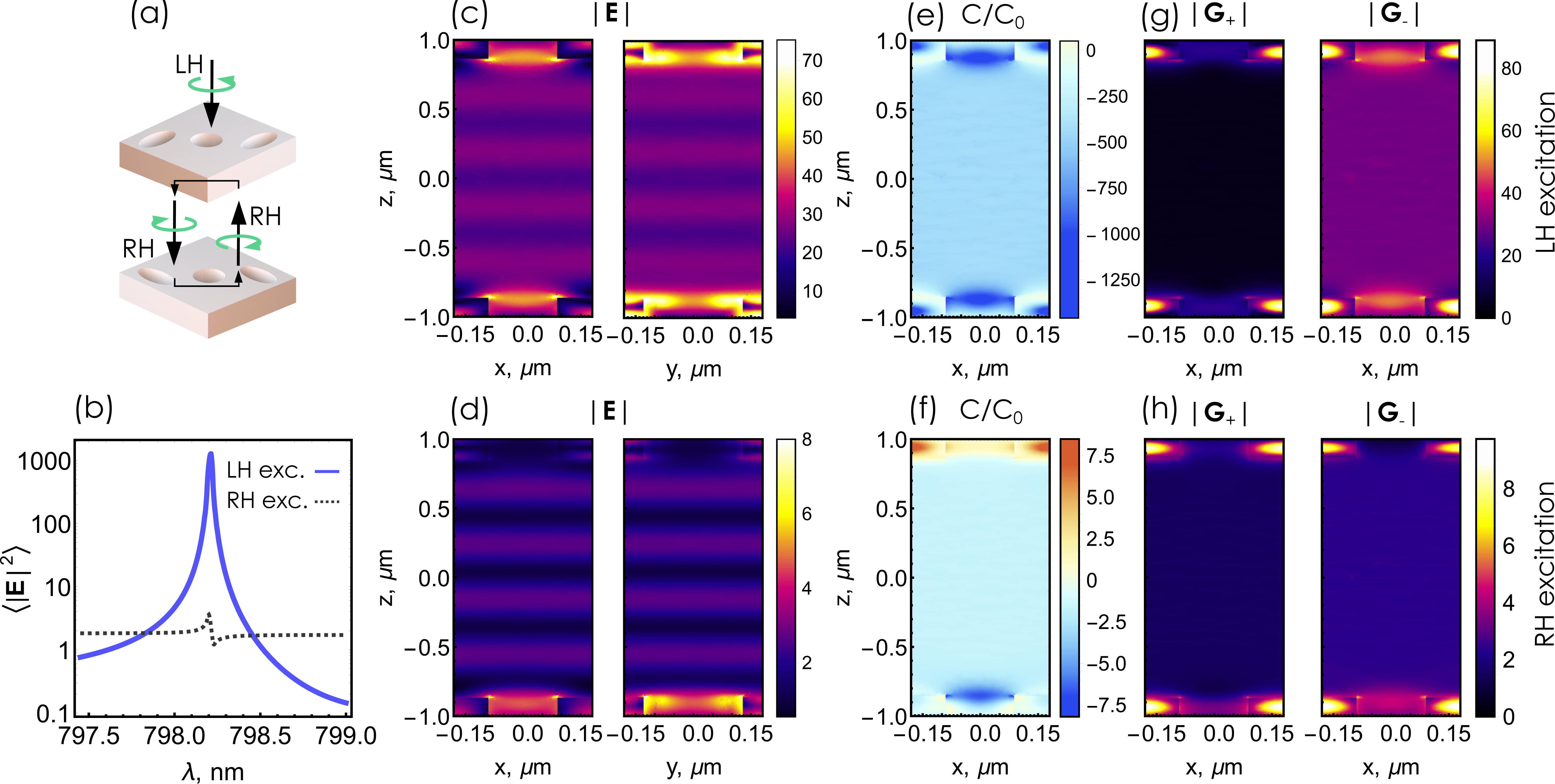}
\caption{FDTD modelling of a single-handedness cavity. (a) Illustration of the formation of a standing chiral wave in the region between two handedness-preserving mirrors. (b) Volume-average electric field enhancement in the region inside the cavity for the cavity thickness $d=1866$ nm for RH and LH excitations at normal incidence. (c) Spatial distribution of electric field intensity inside the cavity induced by the LH incident wave at the resonant wavelength $\lambda_{res}=798.2$ nm; the field is plotted in the two vertical middle planes of the unit cell. (d) The same as (c) for RH excitation. (e) Normalized optical chirality density $C(\mathbf{r})$ for the LH illumination at $\lambda_{res}$ indicating excitation of a pure RH cavity mode. (f) The same as (e) for the RH excitation. (g) Position-dependent RS vector norms $|\mathbf{G}_{\pm}|$ for the LH illumination at $\lambda_{res}$. (h) The same as (g) for the RH excitation.}
\label{fig3}
\end{figure*}

Furthermore, the electric field intensity inside the cavity plotted at $\lambda_{res}$ demonstrates strikingly different spatial distributions for two incident polarizations, Fig. 3(c,d).
Field induced by the incident RH light features a typical standing wave pattern with alternating peaks and troughs caused by the non-zero reflection of the wave by each of the mirrors.
The LH excitation, on the contrary, features a region of nearly uniform electric field intensity. Such a pattern is indicative of the helical standing wave, or so-called polarization standing wave formed by two counter-propagating plane waves of the same handedness \cite{fang2016coherent}.

In agreement with the transfer-matrix calculations, simulated spatial distribution of $C$ at the resonant wavelength $\lambda_{res}$ demonstrates a uniform region of largely enhanced negative chirality density inside the cavity for a LH excitation, Fig. 3(e), suggesting the formation of a RH standing wave.
RH excitation, at the same time, results in a significantly lower magnitude of chirality density, which is also negative, indicating that the RH illumination also partially couples to the RH chiral mode, Fig. 3(f).

To further corroborate the formation of a chiral cavity mode and reveal its pure handedness, we expand the induced field inside the cavity interior into the RH and LH components represented by the corresponding Riemann-Silberstein (RS) vectors \cite{bialynicki2013role}:
\begin{equation}
    \mathbf{G}_{\pm}(\mathbf{r}) = \frac{\mathbf{E} \pm i Z \mathbf{H} }{\sqrt{2}}.
\end{equation}
For any field distribution $[\mathbf{E(r)},\mathbf{H(r)}]$ satisfying Maxwell equations, each of the two RS vectors is an eigenstate of the helicity operator \cite{Fernandez-Corbaton2016}, meaning that it contains only waves of one handedness. According to our handedness convention, $\mathbf{G}_+$ ($\mathbf{G}_-$) represents the LH (RH) field component.

Plots of the position-dependent vector norm $|\mathbf{G}_{\pm}|$, Fig. 3(g,h), clearly show that the LH excitation induces an enhanced RH component between the mirrors (consistent with the fact that handedness of an incident wave flips after transmission through the first mirror), confirming that an eigenmode of a well-defined (right) handedness is excited in the cavity.
RH excitation, on the other hand, only weakly couples to the RH chiral eigenmode owing to the non-zero $t_{RR}$ transmission coefficient. It also induces a non-zero LH field component in the region between the mirrors, originating from the non-resonantly transmitted incident wave ($|t_{LR}| \approx 1$).

We note that although we present in Fig. 3 the solution of the scattering problem and the corresponding total scattered field, the very high Q-factor of the cavity ($>1000$, see Fig. 3(b)) determines that the total field is strongly dominated by the field of the quasi-normal mode of the resonator \cite{lalanne2018light}. Thus, our conclusions inferred from the total field distributions remain valid for the true quasi-normal mode of the chiral resonator.



Next, we examine the possibility to couple a chiral field source to such a single-handedness cavity. A chiral emitter by definition is a pair of collinear electric and magnetic dipoles $\{\mathbf{p},\mathbf{m}\}$ with a $\pm \pi/2$ phase difference between the two, with each phase difference defining one enantiomer state of the chiral source \cite{Shao1997,Klimov2012}:
\begin{equation}
    \mathbf{m}=\pm i \xi c \mathbf{p},
\end{equation}
where $c$ is the speed of light, and $\xi$ is a real-valued parameter. A linear combination with the `$-$' sign defines an LH enantiomer, while the `$+$' sign defines an RH enantiomer. The value $\xi=1$ corresponds to a perfect chiral dipole which produces electromagnetic fields of the corresponding handedness only \cite{zambrana2016tailoring}. 

Single sources typically couple efficiently with discrete eigenmodes of localized cavities, such as plasmonic nanocavities. Our structure presents a periodic translationally invariant resonator, and effective coupling of emitters with such systems requires either sub-diffraction mode area, or a large group index at the emitter frequency \cite{lodahl2015interfacing}. To demonstrate an enantiomer-selective interaction strength, we model emission of a periodic array of chiral sources by imposing periodic boundary conditions. This creates an array of sources that predominantly couples to $\mathbf{k}_{||}=0$ ($\mathbf{k}_{||}$ being the in-plane wave vector) mode of the cavity.

We place a point chiral dipole oscillating at the resonant frequency $\omega_{res}=2\pi c/\lambda_{res}$ in the center of the unit cell of the cavity as sketched in Fig. 4(a) and simulate the field distribution induced by RH and LH emitters under periodic boundary conditions. 
The resulting simulated electric field distributions, Fig. 4, clearly demonstrate that an emitting array of the matching (right) handedness interacts with the cavity mode more efficiently and produces stronger fields, compared to an array of the opposite (left) handedness, that barely interacts with the cavity and produces much weaker fields.

\begin{figure*}[h!]
\includegraphics[width=.9\textwidth]{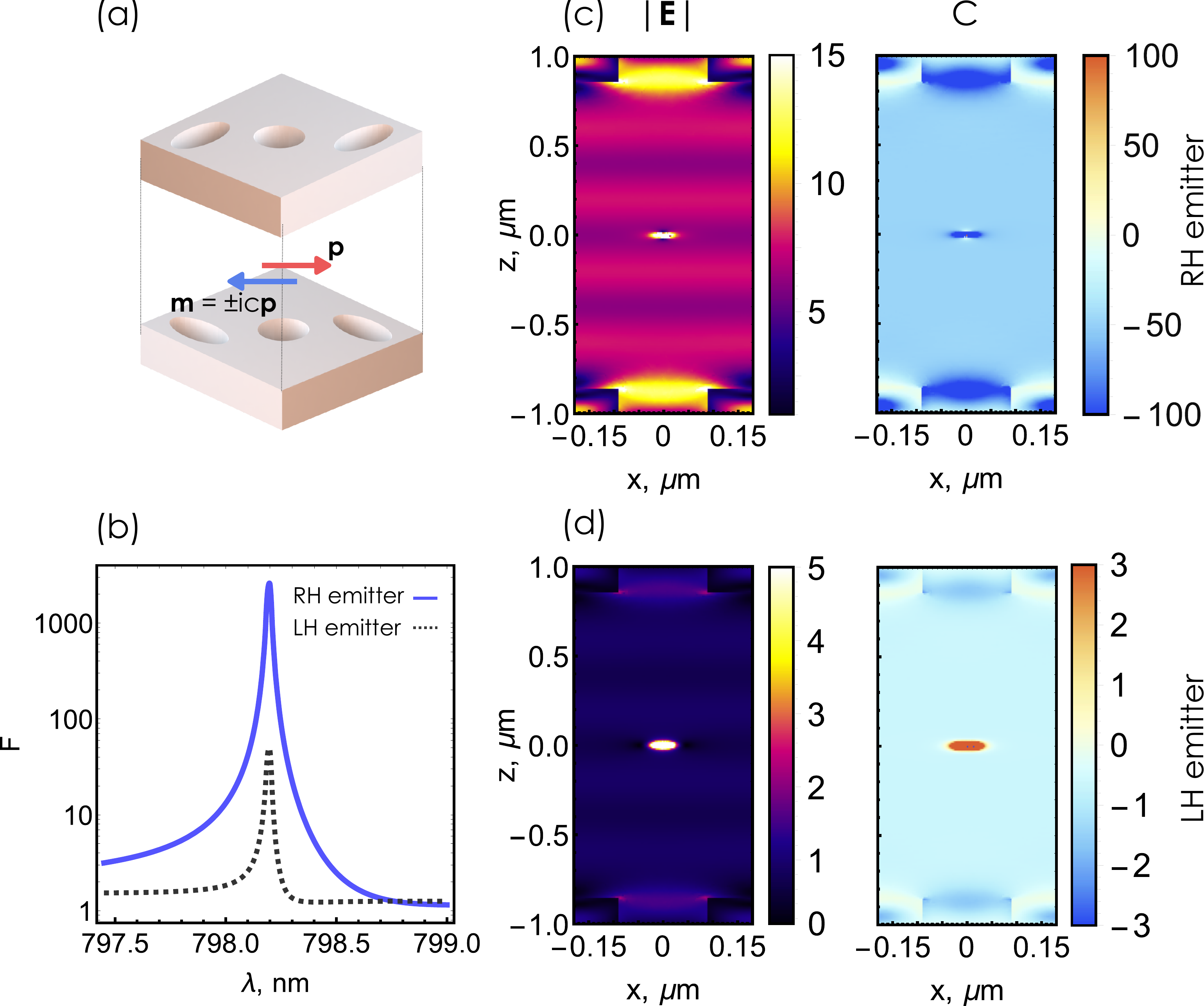}
\caption{Coupling of chiral sources to the chiral eigenmode. (a) Illustration of the chiral cavity excited by a periodic array of point chiral dipoles positioned in the middle of the unit cell. (b) SE enhancement factor $F$ for a array of chiral dipoles, each at the center of a unit cell, for the cavity thickness $d = 1866$ nm. (c) Spatial distribution of the electric field intensity inside the cavity produced by an array of RH chiral emitters at the resonant wavelength $\lambda_{res}=798.2$ nm; the corresponding chirality density distribution. (d) The same as (c) for an array of LH chiral sources.}
\label{fig4}
\end{figure*}

We quantify the degree of the emitter-cavity coupling by the spontaneous emission (SE) rate enhancement factor $F$ (the Purcell factor). The SE enhancement factor is obtained by calculating the emission intensity $P$ produced by a single unit cell of the chiral dipoles array inside the cavity, and normalizing it by the emission intensity of the same array $P_0$ in free space:
\begin{equation}
    F = \frac{P}{P_0} ,
\end{equation}
where the emission intensities are obtained by integrating the total Poynting vector $\mathbf{S}$ across the boundary $\partial V$ of the volume enclosing the chiral dipole: $P = \oint _{\partial V} {\mathbf{S} \cdot d\mathbf{a}}$.
The resulting SE rate enhancement factors obtained for the $1866$ nm thick cavity (see Fig. 4(b)) demonstrate an order of magnitude difference between the RH and LH enantiomer states of the dipolar emitter.

To confirm that the array of emitters indeed couples to the RH cavity mode, we plot chirality densities $C$ induced by the RH and LH emitting arrays, Fig. 4(c,d). The resulting maps reveal nearly homogeneous distributions of negative chirality density in the cavity interior for both handednesses of the emitting array. 
Notice a region of positive $C$ appearing around the LH emitter, Fig. 4(d), where the total field is dominated by the primary (non-scattered) field of the LH dipole carrying a positive chirality density. Spatial distributions of the RS vector norms $|\mathbf{G}_{\pm}|$ further confirm excitation of the RH chiral mode by emitters of either handedness, Fig. S10.

Generally, spontaneous emission rate of a dipolar source is determined by the vacuum fluctuations of the electromagnetic field in its environment \cite{Barron2004,lodahl2015interfacing}.
For a monochromatic chiral dipolar emitter oscillating at a frequency $\omega$ located at $\mathbf{r}_0$ the emission intensity can be calculated as \cite{baranov2017modifying}:
\begin{equation}
    P =  \frac{\omega}{2}
   \Im \left[ \mathbf{p}^* \cdot \mathbf{E}(\br_0) + \mathbf{m}^* \cdot \mathbf{B}(\br_0) \right],
\end{equation}
where $\mathbf{E}(\mathbf{r}_0)$ and $\mathbf{B}(\mathbf{r}_0)$ are the electric and magnetic fields generated by the chiral dipole at its position.
Expressing the generated electric and magnetic fields via the electric and magnetic Green's tensors $\dyadG_{e}$ and $\dyadG_{m}$ of the given structure, we rewrite the emission intensity as
\begin{multline}
    P =  \frac{\mu_0 \omega^3}{2} \left(
    |\mathbf{p}|^2  \un  \cdot  \Im 
    \dyadG_{e} \left({\br}_0,{\br}_0;\omega \right)  \cdot \un  +
    \frac{|\mathbf{m}|^2}{c^2} \un  \cdot \Im  
    \dyadG_{m} \left({\br}_0,{\br}_0;\omega \right)  \cdot \un  \right)
    + \\
    \frac{\mu_0 \omega^3}{2} \left(
    \un \cdot \Im \left[ \frac{pm^*}{c} \frac{1}{ik} \nabla \times \dyadG_e({\br},{\br}_0;\omega) \right]\cdot \un - 
     \un \cdot \Im \left[\frac{p^*m}{c} \frac{1}{ik} \nabla \times \dyadG_m({\br},{\br}_0;\omega) \right]\cdot \un
    \right),
\end{multline}
where $\un  = \mathbf{p}/p$ is the unit vector in the direction of the electric dipole of the chiral source, and $\mathbf{m} = m \un$.
The first two terms involving the imaginary parts of $\dyadG_e$ and $\dyadG_m$ at the emitter's position can be recognized as emission intensities of bare electric, $P_{el}$, and magnetic, $P_{m},$ dipoles, respectively \cite{carminati2015electromagnetic,baranov2017modifying}.
The two other terms appear as the result of interference between the electric and magnetic fields produced by the chiral source. Recalling that $m=\pm i \xi c p$ for a chiral point source, we can group the last two terms in Eq. (10) and obtain:
\begin{equation}
    P =  P_{el}+ P_{m} \pm
    \frac{\mu_0 \omega^2c |\mathbf{p}|^2 \xi}{2} 
    \un \cdot \nabla \times \left(
    \Im \dyadG_{e} ({\br}_0,{\br}_0;\omega) + \Im \dyadG_{m} ({\br}_0,{\br}_0;\omega) 
    \right) \cdot \un,
\end{equation}
where the plus and minus sign applies for left and right enantiomers of the point source, respectively.
It is the last term in Eq. (11) that expresses the interference between the electric and magnetic contributions to the total emission rate.
The parameter $\xi$ expresses the chirality of matter, while the factor $\un \cdot \nabla \times \left(   ... \right) \cdot \un$ can be interpreted as the \emph{chirality density of vacuum fluctuations}. 
For free space $ \nabla \times \Im \dyadG_{e,m} (\br_0,\br_0;\omega)=0$, expressing the fact that the emission intensity of two enantiomers of a chiral source is identical in free space.

The observed strong asymmetry in the SE rate enhancement of left and right chiral point sources clearly indicates that the fluctuating electromagnetic field within such a single-handedness cavity acquires a strong chiral character, which can be utilized for engineering of chiral polaritonic states \cite{Baranov2020} and tailoring the enantiomer-specific optical vacua.
In particular, we can envision that by engineering a proper single-handedness cavity it might be possible to realize coupled polaritonic states involving, say, right-handed enantiomer of a given organic substance, while leaving the left-handed enantiomers in the weak coupling regime, which may affect their chemical kinetics in a different way \cite{thomas2019tilting}.

The study of truly chiral polaritons would, however, require an extension of the existing Hamiltonian models, such as the Hopfield Hamiltonian \cite{hopfield1958theory}. One would have to properly include the magnetic degrees of freedom in both the matter and photonic parts of the system, and to describe the interaction between them. To the best of our knowledge, such a model has not been reported yet and will be a subject of future work.




\section{Conclusions}
To conclude, we have developed the concept of a single-handedness chiral optical cavity.
By utilizing the design of handedness-preserving photonic crystal mirrors, we engineered a Fabry-Perot cavity that supports an optical eigenmode of a well-defined handedness, and does not support eigenmodes of the opposite handedness in the relevant wavelength range. Upon excitation, such a cavity features a uniform region of enhanced electromagnetic field intensity and chirality density.
Furthermore, we demonstrated that chiral point sources couple to such a cavity mode with different coupling strengths depending on the handedness of the source, 
enabling formation of chiral polaritonic states with enantiomer-specific vacuum Rabi splitting.
The concept of single-handedness optical cavity adds a new tool for investigations of chiral matter and opens the door towards studies of chiral electromagnetic vacuum states.

\begin{acknowledgement}
The work was supported by the Russian Science Foundation (21-72-00051) and partly by Swedish Research Council (VR Miljö grant 2016-06059). D.G.B. acknowledges support from the Russian Federation President Grant (MK-1211.2021.1.2). The work of M.V.G. was supported by the Ministry of Science and Higher Education of the Russian Federation within the State assignment of FSRC ``Crystallography and Photonics'' RAS.
\end{acknowledgement}

\begin{suppinfo}
  The supporting information contains additional data presenting the results of FDTD simulations, the CMT describing the handedness-preserving mirrors, and the transfer-matrix calculations of the cavities composed of such mirrors.
\end{suppinfo}

\bibliography{chirality}

\end{document}